# THIRD COMPONENTS WITH ELLIPTICAL ORBITS IN THE ECLIPSING BINARIES: EQ TAU, IR CAS, IV CAS, RY AQR & RZ COM

Tvardovskyi D.E.

[1] Odessa I. I. Mechnikov National University, Odessa, Ukraine

[2] Department "Mathematics, Physics and Astronomy", Odessa National Maritime University, Odessa, Ukraine

**Abstract.** This research is our forth article related to the topic of cyclic O-C changes, third components as the physical process that cause these changes and elliptical orbits of the third components. Here five more eclipsing binary stars were investigated: EQ Tau, IR Cas, IV Cas, RY Aqr and RZ Com. All of them have cyclic O-C curve with superposition of parabolic trend. We computed the mass transfer rate, minima possible mass of the third component and their errors for each of the researched stars.

All of the researched stars are well-known eclipsing binaries which were observed during long period of time. Thus, a lot of photometric, photoelectric observations were done by amateur astronomers and by specialized telescopes. All available data from databases AAVSO [1] and BRNO [2] was used in this research as well as results of the previous investigations made by other authors. Firstly, we took some important general parameters from General Catalogue of Variable Stars (GCVS, [3]) and other researches.

Table 1. Some parameters of the studied eclipsing binaries

| Stellar systems | Initial epoch (JD-2400000) | Period (days) | $M_1, M_\odot$ | $M_2, M_\odot$ | Reference |
|---|---|---|---|---|---|
| EQ Tau | 40213.325 | 0.34134848 | 1.22 | 0.539 | [4] |
| IR Cas | 56521.639 | 0.680685 | 1.43 | 1.22 | [5] |
| IV Cas | 56507.365 | 0.998507 | 2.6 | 1.24 | [6] |
| RY Aqr | 52500.021 | 1.966577 | 1.27 | 0.26 | [7] |
| RZ Com | 34837.4198 | 0.33850604 | 1.03 | 0.45 | [8] |

For any of researched stars errors of the masses were not computed. Thus, they were estimated as 7% of the stellar masses, because it is average value of the errors.

Secondly, all previous articles and abstracts were analyzed. Here is a brief overview of previously published results together with our O-C curves. On all figures pink dots are BRNO observations, blue ones are moments of minima which were computed using AAVSO data. Black line is approximation, in addition the ±σ and ±2σ confidence intervals are shown, where σ is an unbiased estimate or the r.m.s. deviation of the points from the fit.

**EQ Tau**

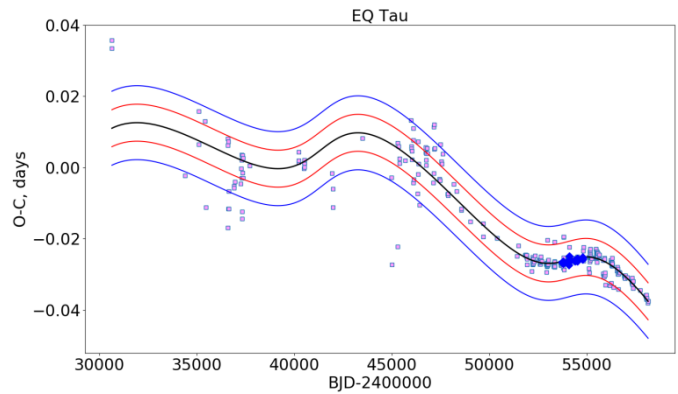

Fig. 1 O-C curve of EQ Tau

Mass transfer rate was computed in [5]. Third component mass was calculated [5] and [9].

**IR Cas**

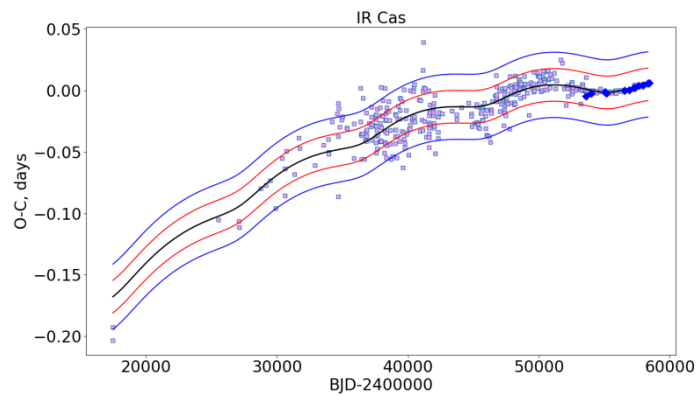

Fig. 2 O-C curve of IR Cas

Mass transfer rate and mass of the third component were computed in [5].

**IV Cas**

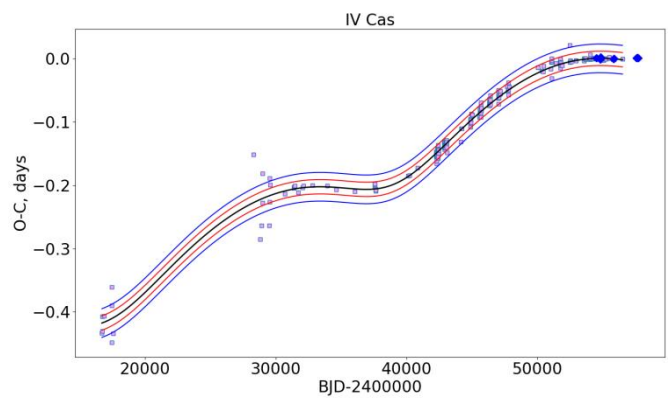

Fig. 3 O-C curve of IV Cas

In [6], [10] and [11] minimal possible mass of the third component was calculated. In [10] and [11] orbital elements were also estimated.

**RY Aqr**

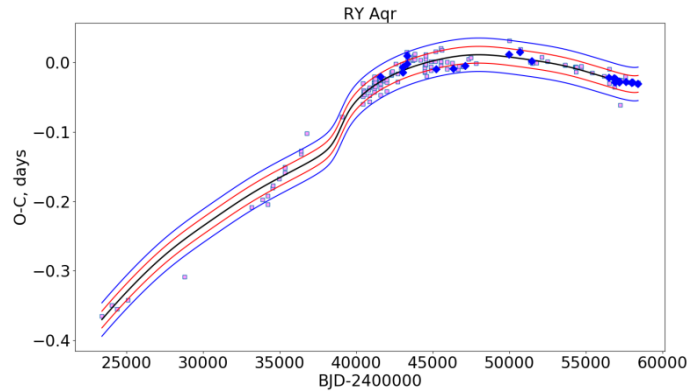

Fig. 4 O-C curve of RY Aqr

In [12] mass transfer rate was estimated and third component's orbital elements were computed. In [7], [12] and [13] third component mass was calculated.

**RZ Com**

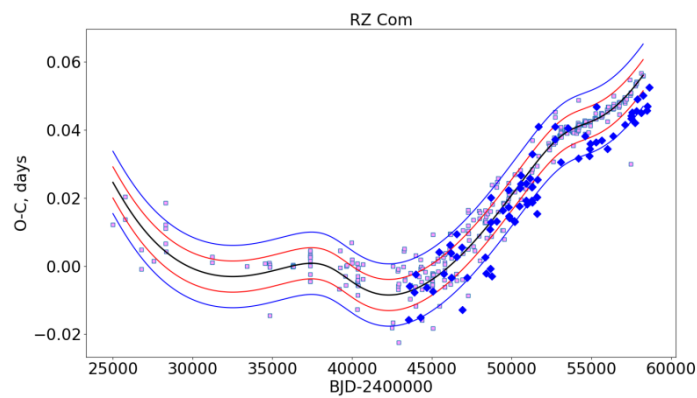

Fig. 5 O-C curve of RZ Com

In [8] mass transfer was supposed and in [14] mass transfer rate was computed. Third component was supposed in [15], [16] and [14], but it's mass was estimated in [15] and [14].

Now it is necessary to describe general aspects of the processing algorithm:

1. Collecting data from database BRNO;
2. Downloading observations from AAVSO;
3. Splitting AAVSO data onto separate minima;
4. Obtaining moment of extremum for each minimum;
5. Joining data form BRNO and obtaining moments of minima;
6. Obtaining values of O-C;

7. Plotting and approximating O-C curves;
8. Obtaining period of cyclic and rate of stable O-C changes;
9. Computing parameters of the physical processes that cause such changes.

For calculating moments of minima from AAVSO observations the software MAVKA was actively used. This code was kindly provided by K.D. Andrych and I.L. Andronov [17], [18], [19]. As the result, 171 minima were obtained.

Table 2. Results of calculations and O-C approximation parameters.

| Value | EQ Tau | IR Cas | IV Cas | RY Aqr | RZ Com |
|---|---|---|---|---|---|
| $\alpha, 10^{-12} \frac{1}{days}$ | -109 ± 4 | -115 ± 8 | -19 ± 13 | -452 ± 15 | 170 ± 2 |
| $\beta, 10^{-6}$ | 8.0 ± 0.4 | 12.8 ± 0.7 | 10.9 ± 0.9 | 48.3 ± 1.3 | -13.0 ± 0.1 |
| $\gamma, days$ | -0.138 ± 0.011 | -0.353 ± 0.016 | -0.566 ± 0.014 | -1.290 ± 0.025 | 0.243 ± 0.003 |
| $a \sin i, 10^6\ km$ | 202 ± 4 | 114 ± 10 | 970 ± 50 | 1300 ± 200 | 177 ± 2 |
| $e, 1$ | 0.31 ± 0.02 | 0.25 ± 0.13 | 0.22 ± 0.04 | 0.82 ± 0.06 | 0.34 ± 0.02 |
| $\omega, rad$ | 3.14 ± 0.06 | 1.9 ± 0.6 | 4.55 ± 0.15 | 6.05 ± 0.04 | 5.58 ± 0.08 |
| $t_0, MJD$ | 9700 ± 500 | 4000 ± 2000 | 16800 ± 1700 | 18700 ± 900 | 16600 ± 400 |
| $T, days$ | 12700 ± 100 | 9400 ± 300 | 21900 ± 800 | 20200 ± 365 | 14900 ± 132 |
| $\dot{M}, 10^{-9} \frac{M_\odot}{year}$ | 26±4 | 220±150 | 11±7 | 36±3 | 33±4 |
| $M_3, M_\odot$ | 0.20±0.01 | 0.17±0.02 | 1.25±0.13 | 1.11±0.29 | 0.14±0.01 |

For better visibility, the possible orientations of the orbits (within the errors of all elements) are shown on the Fig. 6. Solid colorful lines are hundreds of orbits with different orbital elements that could describe observed O-C curves. To plot each of these orbits, some orbital elements ($a$, $e$, $\omega$) were evaluated in range (mean value − error; mean value + error) with relatively small step.

Black arrows show direction to the observer. Black "+" sign in the center of each subfigure is the position of the triple system barycenter. Units of measurement on the axis are million kilometers.

### Acknowledgements

This research was done as the part of project was done as the part of the projects Inter-Longitude Astronomy [20], [21], UkrVO [22], [23] and AstroInformatics [24], [25] as well as previous researches [26], [27], [28].

We sincerely thank to Ivan L. Andronov for fruitful discussions and to AAVSO and BRNO databases for providing data for this research. In addition, we are grateful to Kateryna

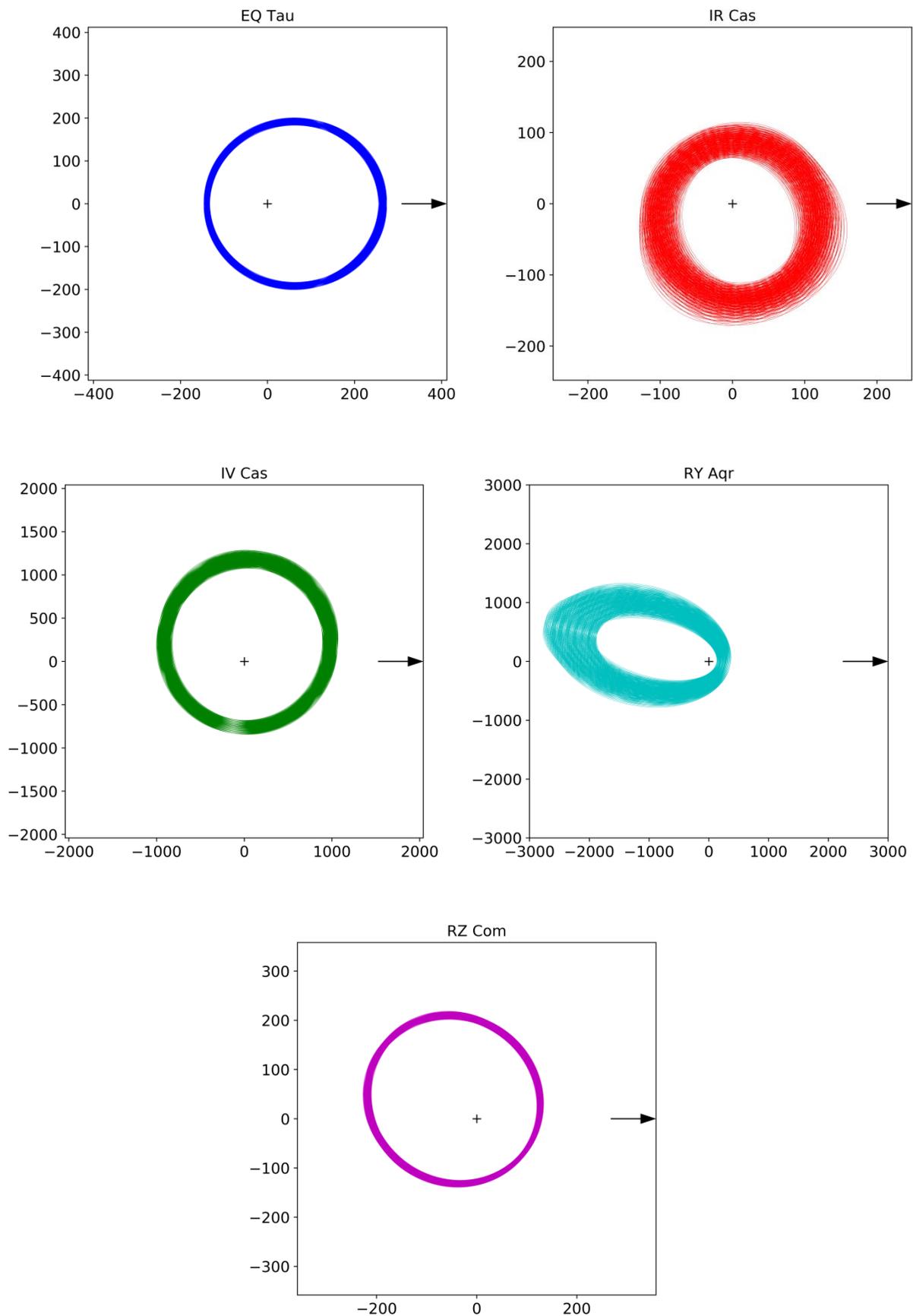

Fig 6. Possible orientations of the binary systems' orbits and ±1σ "error corridors"